\documentclass[a4paper,11pt]{article}

\usepackage{jinstpub}
\usepackage{titlesec}

\usepackage{jinstpub}
\usepackage{lineno}

\usepackage[T1]{fontenc}
\usepackage{enumitem}
\usepackage{soul}
\usepackage{textcomp}
\usepackage{textgreek}
\usepackage[utf8]{inputenc}
\usepackage{booktabs}
\usepackage{tabularx}
\usepackage{multirow}
\usepackage{booktabs}
\usepackage{rotating} 
\usepackage[utf8]{inputenc}
\usepackage{rotating} 
\usepackage{tablefootnote}
\usepackage{array}

\usepackage{float}
\usepackage{footnote}          
\makesavenoteenv{tabularx}     
\usepackage{adjustbox}

\usepackage{tabularx}
\newcolumntype{C}{>{\centering\arraybackslash}X}
\usepackage[flushleft]{threeparttable} 
\modulolinenumbers[1]

\usepackage{subfigure}
\usepackage{makecell} 
\usepackage{booktabs}
\usepackage{orcidlink}
\usepackage{comment}

\usepackage{xcolor}
\definecolor{LinkColor}{rgb}{0,0,0.5} 

\title{Investigating Optical Coupling Material Candidates in NaI(Tl) Detectors for the SABRE-North Dark Matter Experiment}

\author[a,b,1]{Chemseddine Ananna
  \orcidlink{0000-0003-2513-7348},%
  \note{Corresponding author.}}

\author[b]{Gabriella Cataldi
  \orcidlink{0000-0001-8066-7718},}

\author[c,d]{Davide D'Angelo
  \orcidlink{0000-0001-9857-8107},}

\author[e]{Giulia D'Imperio
  \orcidlink{0000-0002-2945-0983},}

\author[a,b]{Maria Luisa De Giorgi
  \orcidlink{0000-0002-3774-6477},}

\author[f]{Giuseppe Di Carlo,
  \orcidlink{0000-0002-3430-2934},}
  
\author[f]{Attilio Di Giacinto
  \orcidlink{0009-0003-7580-7685},}

\author[g]{Marcella Diemoz
  \orcidlink{0000-0002-3810-8530},}

\author[f]{Aldo Ianni
  \orcidlink{0000-0002-6962-3682},}

\author[a,b]{Sana Gul Khattak
  \orcidlink{0009-0007-9122-4219},}

\author[d]{Edoardo Martinenghi
  \orcidlink{0009-0009-9704-5056},}

\author[b]{Alessandro Miccoli
  \orcidlink{0000-0001-7279-504X},}

\author[h]{Marcin Misiaszek
  \orcidlink{0000-0001-5726-9666},}

\author[a,b]{Daniele Montanino
  \orcidlink{0000-0003-2669-1340},}

\author[e]{Valerio Pettinacci
  \orcidlink{0000-0001-8216-7282},}

\author[f]{Lidio Pietrofaccia,}

\author[e,g]{Shahram Rahatlou
  \orcidlink{0000-0001-9794-3360},}

\author[f]{Krzysztof Szczepaniec
  \orcidlink{0000-0003-3229-777X},}

\author[e]{Claudia Tomei
  \orcidlink{0000-0002-7549-6229},}

\author[c,d]{Valerio Toso
  \orcidlink{0000-0002-4853-4613},}

\author[f]{Chiara Vignoli
  \orcidlink{0000-0002-8470-2389},}

\author[b,i]{Samina Zuhra
  \orcidlink{0000-0002-7037-6770},}

\author[f]{Francesco Ferella
  \orcidlink{0000-0003-4264-3170},}

\author[f]{and Stefano Nisi}

\affiliation[a]{
Dipartimento di Matematica e Fisica ``Ennio De Giorgi'',
Università del Salento,\\
Via Arnesano, 73100 Lecce, Italy}

\affiliation[b]{
Istituto Nazionale di Fisica Nucleare, Sezione di Lecce,\\
Via Arnesano, 73100 Lecce, Italy}

\affiliation[c]{
Dipartimento di Fisica ``Aldo Pontremoli'',
Università degli Studi di Milano,\\
Via G. Celoria 16, 20133 Milano, Italy}

\affiliation[d]{
Istituto Nazionale di Fisica Nucleare, Sezione di Milano,\\
Via G. Celoria 16, 20133 Milano, Italy}

\affiliation[e]{
Istituto Nazionale di Fisica Nucleare, Sezione di Roma,\\
Piazzale Aldo Moro 2, 00185 Roma, Italy}

\affiliation[f]{
Istituto Nazionale di Fisica Nucleare,
Laboratori Nazionali del Gran Sasso,\\
Via Giovanni Acitelli 22, 67100 Assergi (AQ), Italy}

\affiliation[g]{
Dipartimento di Fisica, Sapienza Università di Roma,\\
Piazzale Aldo Moro 2, 00185 Roma, Italy}

\affiliation[h]{
Jagiellonian University,
M. Smoluchowski Institute of Physics,\\
Lojasiewicza 11, 30-348 Kraków, Poland}

\affiliation[i]{
Dipartimento di Fisica e Astronomia ``Galileo Galilei'',
Università degli Studi di Padova,\\
Via F. Marzolo 8, 35131 Padova, Italy}

\emailAdd{chemseddine.ananna@le.infn.it}

\abstract{Direct dark matter detection with sodium iodide (NaI) based scintillator detectors requires the use of ultra-low radioactive background materials. Silicone optical pads may serve as the coupling medium between the NaI(Tl) crystal and the Photomultiplier Tubes (PMTs), provided they satisfy stringent requirements in both optical performance and radiopurity. This work reports the characterization of four candidate materials for the SABRE-North experiment. One is the commercial EJ-560 silicone pad, used as a reference; the other three —SYLGARD 184, EGel 3002 A/B, and a hybrid material— were produced locally at the University of Salento under varying catalyst ratios and curing conditions. In total, these four materials yielded ten samples. Optical transmittance was measured over the 310--850\,nm range using a UV/Vis/NIR spectrophotometer, while radiopurity screening was performed  at the STELLA facility (LNGS) using high-purity germanium (HPGe) $\gamma$-ray spectrometry and complemented by high-resolution ICP-MS for the EJ-560 reference. 
The EJ-560 pads achieve an absorbance coefficient of 0.064 $\pm$ 0.017 at the 410\,nm NaI(Tl) emission peak, while the best locally produced samples reached 0.097 $\pm$ 0.015 for the 10:2 base-to-catalyst ratio cured at room temperature, 0.093 $\pm$ 0.014 for that cured at 40$^\circ$C, 0.125 $\pm$ 0.015 for the 2B:A ratio EGel 3002 also cured at 40$^\circ$C, and 0.106 $\pm$ 0.014 mm\textsuperscript{-1} for the hybrid 10:1 EGel\,A:CataSYL formulation cured at room temperature. 
Radiopurity measurements reveal complementary contamination profiles: EJ-560 exhibits detectable $^{232}$Th-chain activity at the $\sim$1.8\,mBq/kg level and the highest $^{40}$K concentration at $(120\pm20)$\,mBq/kg, while the locally produced materials show no detectable Th-chain  contamination but carry $^{226}$Ra activities of $\sim$2.3\,mBq/kg and lower $^{40}$K levels. No single material is found to be optimal across all criteria. Further characterization — including ICP-MS screening of the locally produced compounds, HPGe measurements of cured pads, and direct light-yield tests in a NaI(Tl)--PMT assembly — is identified as necessary to reach a conclusive material selection for the SABRE-North detector.}


\begin{document}
\maketitle
\raggedbottom

\section{\label{sec:intro}Introduction}

The hypothetical Weakly Interacting Massive Particles (WIMPs) form of dark matter is expected to have a very low interaction cross section with hadronic matter. The now decommissioned DAMA/LIBRA experiment claims a WIMP modulation amplitude of the single-hit scintillation events of (0.0116 ± 0.0013) cpd/kg/keV in NaI-based detectors. The internal background of the detectors becomes one of the main challenges in such rare events experiments since cosmic and environmental backgrounds are heavily suppressed in shielded setups located at the deep-underground facility of Laboratori Nazionali del Gran Sasso (LNGS) where the DAMA/LIBRA experiment took place~\cite{Campostrini2024}\cite{best2016}\cite{AHLEN1990}\cite{ANANNA2024}. The SABRE-North experiment -also set at LNGS - aims to independently test DAMA/LIBRA results by producing higher purity crystals for an overall lower internal background. Its Proof-of-Principle (PoP) measurements, taken in 2020~\cite{Calaprice2021} , reached a background level of 1 cpd/kg/keV, which is comparable to that of DAMA/LIBRA in the relevant 2 to 6 keV energy region of interest (ROI), where the WIMP-compatible modulation has been observed. However, for its final setup, SABRE-North aims for a further 50\% reduction in the low energy ROI count rate through the use of zone refined NaI powder to grow the scintillator crystals \cite{ananna2026}. 

Scintillation light collection is also crucial aspect in investigating low energy events since the contribution of each single photon has a significant impact on the measured signals. The optical coupling material used between the NaI crystal and the Photomultiplier Tubes (PMT) glass window must therefore maximize  scintillation light transmittance while having a limited contribution to the overall radioactive background. This work investigates three candidates for optical coupling pads material for a possible use in the SABRE-North experiment detectors: SYLGARD Silicone 184, EGel 3002 A/B, EJ-560 silicone. The EJ-560 silicone pads are directly sourced from the manufacturer while SYLGARD and EGel pads are produced locally in the optics laboratory at the University of Salento - Italy (UniSalento). Variations in the production process were also tested with the goal of increasing transmittance, resulting in ten samples in total. The details regarding samples preparation are discussed in section \ref{sec:materials}. The transmittance measurements runs are also carried locally on the 10 + 1 samples (10 locally produced + 1 EJ-560 reference). The methodology and results of these latter are reported in sections \ref{sec:Transmittence_measurements} and \ref{subsec:transmittance_results} respectively.

Potassium-40 and Lead-210 (from Uranium-238) are the main background contributors in the energy ROI of SABRE-North. These elements are found in the optical coupling pads at different amounts depending on the material but may also vary from sample to sample. A radioactive contamination screening of the pads is therefore required. This measurement was performed at LNGS using high-purity Germanium detectors (HPGe) and High Resolution ICP-MS\footnote{Inductively Coupled Plasma-Mass Spectrometry.}. The samples internal contamination of the investigated radioactive isotopes are respectively presented in Section \ref{subsec:screening_results}.

\section{\label{sec:materials} Materials and Sample preparation}

\subsection*{EJ-560 commercial pad (Reference)}

The first material investigated in this work is the EJ-560 silicone rubber, manufactured by Eljen Technology (Sweetwater, TX, USA) \cite{eljen_ej560}. It is a transparent flexible elastomer that slightly deforms when pressed but easily regains its original shape when pressure is relieved. This elastic recovery provides a degree of surface conformability and sealing, complementing its primary purpose as an optical coupling medium. EJ-560 pads have been widely adopted as the optical coupling choice in low-background experiments where maximizing light collection efficiency is critical, despite their relatively high cost \cite{Choi2024}\cite{Lee2023}\cite{Moore2025}. The EJ-560 silicone is commercially provided either in large 1 to 3 mm thickness sheets or in the form of precut circular pads. For this study, square samples of $\sim$ 26.5 mm side length were cut from a larger sheet as supplied by the manufacturer and used without further processing or surface treatment. The sample thickness was measured to be 1.22 ± 0.06 mm (N = 10 samples), indicating good thickness uniformity across the batch (see Table \ref{tab:thick_transmittance}). According to the manufacturer's datasheet, the EJ-560 material is claimed to have $\>$90\% bulk light transmittance near the NaI(Tl) scintillation emission peak ($\sim$410 nm) with a refractive index of 1.43 at at the standard sodium D-line (589 nm). Additional physical and mechanical properties are summarized in Table \ref{tab:SiliconeComparison}. The EJ-560 samples serve as a reference material against which the optical transmittance and radioactive contamination screening results of the locally produced, more cost-effective SYLGARD 184 and EGel 3002 pads are benchmarked.

\subsection*{SYLGARD 184 Silicone Elastomer}

The second material investigated is the \textbf{SYLGARD 184 Silicone Elastomer}, a polydimethylsiloxane-based (PDMS) two-part silicone manufactured by Dow Inc. (Midland, MI, USA). It is primarily designed for potting and encapsulating electronic components such as printed circuit boards (PCBs), LEDs, and sensors, and also exhibiting flame-retardant properties \cite{Sylgard_184}. It is supplied as a kit consisting of two liquid components: a base polymer solution (we refer to as \textbf{BaseSYL}) and a platinum-based crosslinking catalyst (\textbf{CataSYL}). This nomenclature is adopted for the rest of the work to avoid confusion with the analogously named component solutions of the EGel 3002 system described in Section~\ref{sec:materials}. It is also worth noticing that the manufacturer does not disclose detailed chemical compositions for either component. The recommended fabrication procedure consists of mixing \textbf{BaseSYL} and \textbf{CataSYL} in a 10:1 mass ratio, followed by curing at room temperature (25°C) for 48 hours. Alternatively, curing can be thermally accelerated at 100°C for 35 minutes or at 150°C for 10 minutes. The resulting fully cured material is a colorless, transparent, flexible elastomer with a nominal hardness of approximately 43 Shore A. For this study, 6 samples were produced using varying component ratios and curing conditions, as summarized in Table~\ref{tab:SamplesListingNew}. A consistent mass of 1.2 g of the mixed solution was poured into 26.55 mm square hard plastic molds to identically shape the samples. The curing processes are performed in a Specacabinet™ drying cabinet that is thermostatically controlled up to 80ºC (stability $\pm$0.5ºC). 

Preliminary attempts using PTFE molds resulted in degraded sample uniformity and optical transparency, bubble formation -no outgassing used for this work-, and difficulty releasing the cured pad from the mold; hard plastic molds were therefore selected for all subsequent sample production. Although sample hardness was not directly measured in this work, the mechanical feel of the produced pads was qualitatively consistent with that of the EJ-560 commercial silicone optical pads. The \textbf{CataSYL} solution was also later used as the crosslinking agent in the fabrication of two additional hybrid samples incorporating the EGel 3002 \textbf{A}-component, as reported in Section~\ref{sec:materials}.

\subsection*{EGel 3002 A/B}

Similarly, \textbf{EGel 3002 A/B} is a tacky silicone elastomeric gel used to protect electronic components. It is also prepared by mixing two solutions supplied as a kit by \textit{CHT Silicones}.
The manufacturer does not disclose the exact chemical composition of either solution; 
however, it is known that the \textbf{A}-component (i.e. the catalyst) contains trace amounts of platinum \cite{VIBA2019}. The second solution contains the \textbf{B}-component (or \textbf{B}) which serves as the base of the mixture. A straightforward 1:1 mass ratio of the two components is recommended by the manufacturer, with curing maximum time and temperature conditions of 24h at 25°C or 30~min at 100°C. 

These standard settings are, however, intended for the production of a very adhesive, gel-like substance that we found to be easily destroyed upon peeling from a rigid mold. Variations from the recommended settings are therefore investigated to produce firmer pads similar to the previous two materials (i.e. EJ-560 and Sylgard 184). We achieved the targeted consistency by doubling the relative amount of the \textbf{B}-component. Different curing temperature and duration combinations have also been investigated using the same Specacabinet™ drying cabinet, with the most promising results reported in Table \ref{tab:SamplesListingNew}. Pad samples \#9 and \#10 are from a hybrid material we produced using the \textbf{CataSYL} solution as the catalyst mixed to the EGel 3002 \textbf{A}-component, used here as the base, with a 10:1 mass ratio. Parallel trials using the \textbf{B}-component as the base alongside the \textbf{CataSYL} catalyst resulted in no polymerization.

\begin{figure}[]
    \centering
    \includegraphics[width=\textwidth]{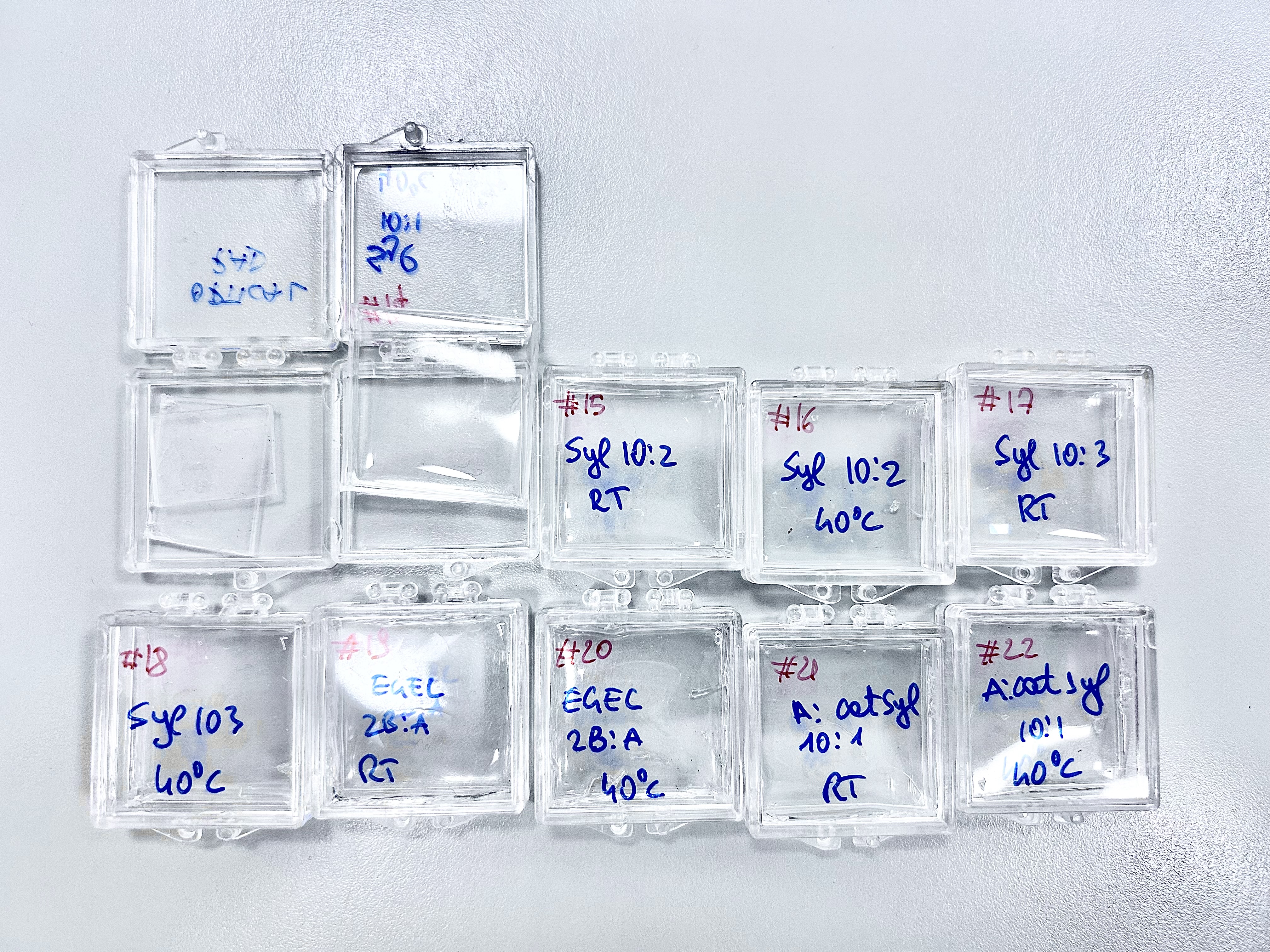}
    \caption[Silicone pads samples]{Silicone pads in their curing plastic molds. Picture taken circa a year after their production date. Container numbers (in red) unrelated to sample numbering used in this work.}
    \label{fig:samples_photo}
\end{figure}

In total, 10 samples are investigated for this work: 6$\times$Sylgard 184-based, 2$\times$EGel 3002 A/B-based, and 2$\times$Sylgard/EGel hybrid pads. All have been optically characterized for their transmittance. ICP-MS radiopurity measurements were only performed on to EJ-560 samples, while HPGe-based screening is directly performed at LNGS on uncured Sylgard 184 and EGel 3002 A/B mixtures as reported in Section \ref{subsubsec:HPGe}. It is also worth mentioning that none of the locally produced samples showed visible signs of photo-oxidation (yellowing) after almost two years of storage under standard ambient conditions (i.e. exposed to air at room temperature, without direct solar UV radiation) (see Figure \ref{fig:samples_photo}).

\begin{table}[ht]
\centering
\begin{threeparttable}
\begin{tabular}{|l|c|c|c|}
\hline
\textbf{Property} & \textbf{EJ-560} & \textbf{Sylgard 184} & \textbf{EGel 3002 A/B} \\ \hline
Density (g/cm$^{3}$) & 1.03 & 1.03 & 0.97 \\ \hline
Hardness (Shore) & 16--24 A & 43--50 A & 40--55 (00) \\ \hline
Refractive Index (589 nm) & 1.43 & 1.41 & 1.40 \\ \hline
Operating Temp.\ Range ($^{\circ}$C) & $-40$ to $70$ & $-45$ to $200$ & $-55$ to $200$ \\ \hline
CTE\tnote{*} (cm/cm$\cdot{}^{\circ}$C) &
$3.0\times10^{-4}$ &
$3.1\times10^{-4}$ &
$3.1\times10^{-4}$\cite{VIBA2019} \\ \hline
\end{tabular}

\begin{tablenotes}
\footnotesize
\item[*] Coefficient of Linear Thermal Expansion.
\end{tablenotes}

\caption{Comparison of the nominal values of some physical properties of EJ-560, Sylgard 184, and EGel 3002 A/B.}
\label{tab:SiliconeComparison}
\end{threeparttable}
\end{table}

\section{Transmittance measurements}
\label{sec:Transmittence_measurements} 

The optical transmittance of locally fabricated SYLGARD 184 and EGel 3002 A/B samples, as well as commercially sourced EJ-560 pads, was characterized using a Lambda 900 UV/Vis/NIR spectrophotometer (PerkinElmer Inc., Waltham, MA, USA). The instrument employs a dual-lamp illumination system comprising a deuterium (D\textsubscript{2}) lamp and a tungsten-halogen lamp. The D\textsubscript{2} lamp provides spectral coverage in the ultraviolet region up to approximately 320 nm, while the tungsten-halogen lamp covers the visible and near-infrared (NIR) regions up to $\sim$900 nm \cite{Lambda900}\cite[p.~151]{Skoog2018}. When the selected wavelength range spans both regions, as in this work, the instrument performs an automatic lamp changeover at $\sim$319 nm without interrupting the scan.
Transmittance was measured by directing a collimated monochromatic beam perpendicularly through samples with thicknesses ranging from 1.22 to 1.62 mm (detailed in Table \ref{tab:thick_transmittance}), with detection carried out by an R6872 photomultiplier tube (PMT), the designated detector for the UV/Vis/NIR range in this instrument. The resulting photocurrent signals were digitized and recorded by the data acquisition (DAQ) system. Transmittance spectra were acquired over a wavelength range of 310–850 nm in 4 nm increments, covering the full region of interest (ROI) relevant to the NaI(Tl) scintillation emission which peaks at approximately 410 nm. Prior to the measurement session, a single baseline correction was performed using an open-beam reference scan to account for ambient background contributions. All measurements were performed consecutively in a single session, following a warm-up period to allow the light source to stabilize. Baseline drift and the associated uncertainty due to source instability are therefore assumed to be negligible.
Measurements of the refractive index \textit{n} were also attempted, exploiting optical phenomena at the interface between transparent media of differing refractive indices, including total internal reflection, Brewster's angle, and Fresnel coefficient analysis. However, the light source instability produced uncertainties that precluded determination of \textit{n} with the precision required for this experiment.

\begin{table}[H]
\centering
\scriptsize
\caption{Listing of the 10 samples produced for this work. The columns represent, in order, the sample number (name), material type, base solution, catalyst solution, their mixing ratio, the quantity of mixture poured into the mold, and the corresponding curing temperature/time.}

\begin{threeparttable}

\makebox[\textwidth][c]{%
\resizebox{1.1\textwidth}{!}{%
\begin{tabular}{l l l l c c c}
\toprule
\textbf{Sample} & \textbf{Type} & \textbf{Base} & \textbf{Catalyst} &
\textbf{Ratio} & \textbf{Quantity (g)} &
\makecell[c]{\textbf{Temperature-}\\\textbf{time}} \\
\midrule

\#1 Syl 10:1 RT       & Sylgard & BaseSYL & CataSYL & 10:1 & 1.2 & RoomT\tnote{*} (48 h) \\
\addlinespace
\#2 Syl 10:1 40C      & Sylgard & BaseSYL & CataSYL & 10:1 & 1.2 & 40$^\circ$C (48 h) \\
\addlinespace
\#3 Syl 10:2 RT       & Sylgard & BaseSYL & CataSYL & 10:2 & 1.2 & RoomT\tnote{*} (48 h) \\
\addlinespace
\#4 Syl 10:2 40C      & Sylgard & BaseSYL & CataSYL & 10:2 & 1.2 & 40$^\circ$C (48 h) \\
\addlinespace
\#5 Syl 10:3 RT       & Sylgard & BaseSYL & CataSYL & 10:3 & 1.2 & RoomT\tnote{*} (48 h) \\
\addlinespace
\#6 Syl 10:3 40C      & Sylgard & BaseSYL & CataSYL & 10:3 & 1.2 & 40$^\circ$C (48 h) \\
\addlinespace
\#7 EGEL 2B:A RT      & EGEL    & B       & A       & 2:1  & 1.2 & RoomT\tnote{*} (48 h) \\
\addlinespace
\#8 EGEL 2B:A 40C     & EGEL    & B       & A       & 2:1  & 1.2 & 40$^\circ$C (48 h) \\
\addlinespace
\#9 A:CataSYL 10:1 RT & Hybrid  & A       & CataSYL & 10:1 & 1.2 & RoomT\tnote{*} (48 h) \\
\addlinespace
\#10 A:CataSYL 10:1 40C & Hybrid & A      & CataSYL & 10:1 & 1.2 & 40$^\circ$C (48 h) \\
\bottomrule
\end{tabular}%
}
}

\begin{tablenotes}
\footnotesize
\item[*] Room temperature; not strictly controlled.
\end{tablenotes}

\end{threeparttable}

\label{tab:SamplesListingNew}
\end{table}

\section{Radiopurity Screening}
\label{subsec:screening}
\subsection{High-Purity Germanium \textgamma-ray Spectrometry}
\label{subsubsec:HPGe}

The $\gamma$-ray spectroscopy measurements relative to this work were conducted at the STELLA (SubTErranean Low Level Assay) facility at LNGS, Italy, located at the depth of 3\,800~m\,w.e. Operating at such depth reduces the cosmic-ray muon intensity by roughly six orders of magnitude compared to the surface. This reduction substantially lowers the in-situ production of radioactive spallation products and removes the dominant source of  background present in above-ground setups~\cite{best2016, AHLEN1990, Laubenstein2004}. This environment makes it possible to push the sensitivity of HPGe $\gamma$-ray spectrometry into the $\mu$Bq/kg range~\cite{Laubenstein2020}. Three p-type coaxial HPGe detectors available at the STELLA facility were used for the characterization of the opical coupling pads: \texttt{GsOr}, \texttt{GePaolo}, and \texttt{GeMPI3}. Their crystal masses are $\sim$2.2 kg, $\sim$2.75 kg, and $\sim$2.3 kg respectively, and relative efficiencies\footnote{Following IEEE Std 325-1971.} of 94.2\%, 113.2\%, and 98.7\%. Further properties regarding the detectors are reported in Table~\ref{tab:STELLA_detectors}, and their respective shielding configurations and geometries are described in~\cite{Laubenstein2017} for \texttt{GsOr} and \texttt{GePaolo}, and in~\cite{budjas2008} for \texttt{GeMPI3}.

\begin{table}[H]
\centering
\small
\renewcommand{\arraystretch}{1.4} 
\caption{Properties and background count rates of the HPGe detectors used for the screening. Data compiled from different sources \cite{ARPESELLA2002}\cite{budjas2008}. }
\noindent\hspace*{-\dimexpr\oddsidemargin+0.in\relax}%
\makebox[\paperwidth][c]{%
\begin{tabular}{l l c c c c c c}
\toprule
\multirow{2}{*}{\textbf{Name}} & 
\multirow{2}{*}{\makecell{\textbf{Detector}\\\textbf{Type}}} & 
\multirow{2}{*}{\makecell{\textbf{Vol.}\\\textbf{(cm$^3$)}}} & 
\multirow{2}{*}{\makecell{\textbf{Rel. Eff.}\\\textbf{(\%)}}} & 
\multicolumn{4}{c}{\textbf{Count Rate (d$^{-1}$ kg$^{-1}$)}} \\
\cmidrule(lr){5-8}
& & & & \textbf{60--2700 keV} & \textbf{352 (U)} & \textbf{583 (Th)} & \textbf{1461 (K)} \\
\midrule
\texttt{GsOr}    & p-type & 414 & 94.2  & 442        & 2          & 0.76       & 4.2 \\
\texttt{GePaolo} & p-type & 518 & 113.2 & 222        & 1.1        & 0.31       & 1.8 \\
\texttt{GeMPI 3} & p-type & 430 & 98.7  & $24\pm1$ & $\sim0.5$ & $\sim1$   & $\sim0.5$ \\
\bottomrule
\end{tabular}%
}
\label{tab:STELLA_detectors}
\end{table}

A practical feature of HPGe $\gamma$-ray spectrometry is that samples do not require chemical processing or destruction prior to measurement; detection efficiencies for arbitrary sample geometries are instead determined via Monte Carlo simulations~\cite{Laubenstein2020, Laubenstein2017}. The three materials characterized in this work required different preparation strategies. The EJ-560 pads, being commercially supplied as solid sheets, were measured directly as received, with a total sample mass of $\sim$175\,g. The EGel 3002\,A/B and Sylgard 184 compounds were each prepared as liquid mixtures from their respective two-part solutions following the mass ratios recommended by the manufacturer (see Section~\ref{sec:materials}), yielding sample masses of $\sim$500\,g each. In both cases the screening was directly performed on the liquid mixtures because of the difficulty to cure such large volumes of material. In general, maximizing the sample mass within a compact volume increases the detection solid angle, improves counting statistics, and lowers the achievable minimum detectable activity~\cite{Laubenstein2020, Heusser2006}. The sample masses and counting live-times for each measurement are reported in Section~\ref{subsec:screening_results}.

\subsection{High Resolution ICP-MS measurements}
\label{subsec:ICP-MS_Methodology}

Alongside the HPGe $\gamma$-ray spectrometry, the EJ-560  was also submitted to high resolution ICP-MS based screening. A large foil of the polymer was solubilized and treated with the ashing procedure in order to remove organic components from the sample. The total burning of the foil is achieved using a NABERTHERM muffle with heating ramp at 650°C for 4 hours. A quartz crucibles with caps is used to avoid dust dispersion. Around 40\% of the foil mass remained inside crucibles after the burning cycle, showing the presence of some additives and/or substance of unknown composition with a melting point that is higher than that of usual organic compounds.
In order to finalize analysis, other test using higher temperature and longer time durations was performed and showed to similar results; since the best way to solubilize organic compounds for  ICP-MS analysis remains the ashing process, the next step was to recover all amount of sample ashed and not ashed using a 10\% HNO\textsubscript{3} ultrapure solution at 80°C for 1 hour in oven. After this procedure, the whole solution was recovered from the crucible and measured using a mass spectrometer.
The ICP-MS characterization is performed using a \textit{Thermo Fisher Element 2} mass spectrometer, equipped with a \textit{Cetac ASX520} auto-sampler. To prevent carryover and minimize memory effects, a 60-second rinse with a 2\% nitric acid ultra-pure (HNO\textsubscript{3}) solution is performed between successive measurements. All measurements were carried out within a clean room environment ISO6 class to minimize environmental contamination risk. The measurement focused on Th/U quantification, mainly determining the amount of stable \textsuperscript{232}Th and \textsuperscript{238}U isotopes. A Standard solution with 10\textsuperscript{-12}g/g of Th and U was used to calibrate the instrument, and the following analysis performed in quantitative mode.
The data acquisition of the U and Th isotopes measurement is performed in low and medium resolution in order to solve possible interferences caused by the presence of organic matrix in sample; in fact with this type of mass spectrometer it’s possible to work in low, medium, high resolution. For this specific case, medium resolution was necessary to determine exact amount of specific isotopes (results shown in table \ref{tab:screening_results}). The acquisition resolution mode (i.e. low, medium, and high) relates the separation of isobaric interferences, as polyatomic species formed by presence of argon, oxygen, chlorine, fluorine, etc. Isobaric interferences arise from the direct overlap of isotopes from different elements that share the same nominal mass, a known limitation of the ICP-MS technique~\cite[p.~265]{Skoog2018}. Low-resolution instruments are unable to resolve these overlaps. However, such interferences can often be minimized by selecting a non-interfering isotope when analyzing multi-isotopic elements or by applying mathematical correction methods.
Polyatomic ions—particularly argides, hydrides, carbides, nitrides, and oxides—are molecular species that form naturally in the argon plasma~\cite{profrock2012}. Their precursors originate from the argon gas itself, as well as from solvents, acids used during sample preparation, and components of the sample matrix.
Several strategies have been developed to reduce, and in some cases completely eliminate, polyatomic interferences. These include the use of high-resolution ICP-MS (HR-ICP-MS) to improve mass resolution, cold plasma conditions for elements with low ionization potential such as K, Na, Ca, and Sr, and the implementation of reaction or collision cells~\cite{NISI2017}.

\section{Results}
\label{sec:results}

\subsection{Optical Transmittance}
\label{subsec:transmittance_results}

\begin{figure}[H]
\centering
\includegraphics[width=0.8\textwidth]{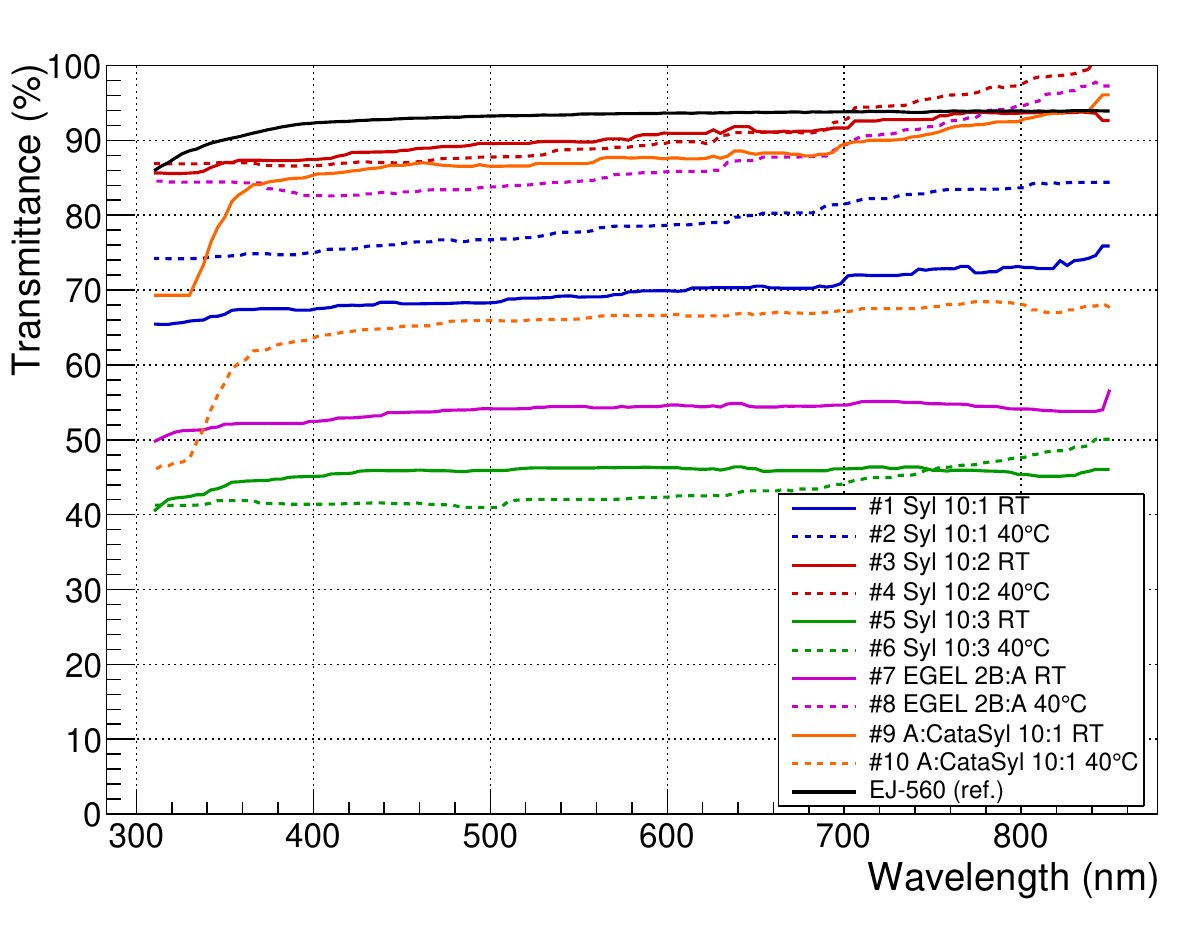}
    \caption{Measured transmittance spectra of the ten locally produced samples and the EJ-560 reference pad over the 310--850\,nm range.
    }
    \label{fig:tranmisttance}
\end{figure}

The measured transmittance values at the 410\,nm NaI(Tl) emission peak of the eleven tested optical coupling pad samples are reported in Table~\ref{tab:thick_transmittance}, grouped by material type: EJ-560 reference (sample~\#0), Sylgard\,184-based (samples \#1 to \#6), EGel\,3002\,A/B-based (samples \#7 and \#8), and hybrid EGel 3002 A-component with Sylgard 184 catalyst (samples \#9 and \#10). Full transmittance spectra over the 310--850\,nm range are shown in Figure~\ref{fig:tranmisttance}. The systematic uncertainty is estimated comparing the measured EJ-560 transmittance curve with that reported in the technical sheet. The two curves diverge by up to 9\% near 330 nm but stay under 2\% above 350 nm — our region of interest (the scintillation emission peak). We therefore adopt 2\% as the relative uncertainty for all subsequent transmittance measurements.

The EJ-560 reference pads achieved 92.5\% transmittance at 410\,nm, consistent with the manufacturer's stated value of $>$90\%~\cite{eljen_ej560}. Among the Sylgard\,184 samples cured at room temperature, transmittance increased from 67.7\% at the standard 10:1 base-to-catalyst ratio (sample~\#1) to 87.6\% at 10:2 (sample~\#3), then decreased to 44.9\% at 10:3 (sample~\#5). The ratio-corresponding pads cured at 40$^{\circ}$C (samples \#2, \#4, \#6) yielded values of 75.5\%, 86.8\%, and 41.4\% respectively, showing no consistent performance trend relative to their room-temperature counterparts. EGel\,3002\,A/B samples produced at the manufacturer-recommended 1:1 ratio did not allow a consistent transmittance measurement due to the highly adhesive, gel-like surface of the resulting pads. The two EGel samples produced at a 2:1 B-to-A ratio reached 52.7\% (room temperature, sample~\#7) and 82.6\% (40$^{\circ}$C, sample~\#8) at 410\,nm. The hybrid A:CataSYL samples (\#9 and \#10), produced at a 10:1 ratio, yielded 85.6\% and 64.0\% respectively. At wavelengths below 350\,nm, most samples exhibit a progressive decrease in transmittance approaching the lower limit of the scanned range, as seen in Figure~\ref{fig:tranmisttance}. Sample thicknesses, measured for each pad, are reported alongside transmittance values in Table~\ref{tab:thick_transmittance}. The locally produced samples are systematically thicker than the EJ-560 reference (1.22\,mm), a consequence of the mold-based casting method used in this work. Since raw transmittance depends on sample thickness, comparing it directly across samples of different thickness may be misleading. The Beer–Lambert law's attenuation coefficient $\mu$ removes this thickness dependence\footnote{from $T = e^{-\mu L}$ where T is the measured transmittance (in \%), $\mu$ is the linear attenuation coefficient, and L the sample thickness.}, making it the better basis for comparison.

\begin{table}
    \centering
    \caption{Transmittance measurements results, calculated attenuation coefficient, and their associated uncertainties. Table columns report in order: samples name, transmittance (T) at 410nm, thickness (L), attenuation coefficient $\mu$, and $\sigma$\textsubscript{$\mu$} and $\sigma$\textsubscript{L} their associated uncertainties within 1$\sigma$.} 
    \label{tab:thick_transmittance}
    \noindent\hspace*{-\dimexpr\oddsidemargin+0.in\relax}%
    \makebox[\paperwidth][c]{%
    \begin{threeparttable}
    \begin{tabularx}{0.95\paperwidth}{l C C C C C}
    \toprule
    \multicolumn{1}{c}{\adjustbox{valign=c}{\textbf{Sample}}} & \textbf{T (\%)} & \textbf{L (mm)} & \textbf{$\sigma$\textsubscript{L} (mm)} & \boldmath{$\mu$ \textbf{(mm\textsuperscript{-1})}} & \boldmath{$\sigma\textsubscript{$\mu$}$ \textbf{(mm\textsuperscript{-1})}} \\
    \midrule
    \#0 EJ-560 (ref.)        & 92.5 & 1.22 & 0.06 & 0.064 & 0.017 \\
    \#1 Syl 10:1 RT          & 67.7 & 1.41 & 0.04 & 0.277 & 0.016 \\
    \#2 Syl 10:1 40C         & 75.5 & 1.58 & 0.04 & 0.178 & 0.013 \\
    \#3 Syl 10:2 RT          & 87.6 & 1.36 & 0.06 & 0.097 & 0.015 \\
    \#4 Syl 10:2 40C         & 86.8 & 1.52 & 0.05 & 0.093 & 0.014 \\
    \#5 Syl 10:3 RT          & 44.9 & 1.59 & 0.08 & 0.504 & 0.028 \\
    \#6 Syl 10:3 40C         & 41.4 & 1.50 & 0.05 & 0.588 & 0.024 \\
    \#7 EGEL 2B:A RT         & 52.7 & $\sim$1.46 &    N/A\tnote{\textasteriskcentered} & $\sim$0.439 &  N/A\tnote{\textasteriskcentered} \\
    \#8 EGEL 2B:A 40C        & 82.6 & 1.53 & 0.09 & 0.125 & 0.015 \\
    \#9 A:CataSYL 10:1 RT    & 85.6 & 1.47 & 0.02 & 0.106 & 0.014 \\
    \#10 A:CataSYL 10:1 40C  & 64.0 & 1.62 & 0.14 & 0.275 & 0.027 \\
    \bottomrule
    \end{tabularx}
    \begin{tablenotes}
        \footnotesize
        \item[\textasteriskcentered]Irregular and sticky pad surfaces impeding to perform a consistent series of measurements.
    \end{tablenotes}
    \end{threeparttable}%
    }
\end{table}

\subsection{Radiopurity screening}
\label{subsec:screening_results}
The results of the HPGe $\gamma$-ray spectrometry screening are summarised in Table~\ref{tab:screening_results}. Sylgard\,184 and EGel\,3002\,A/B liquid samples were measured with detectors \texttt{GsOr} and \texttt{GePaolo} for live-times of 20.4\,d and 34.1\,d, with sample masses of 548.8\,g and 502.2\,g respectively. The EJ-560 solid sheets were measured with \texttt{GeMPI3} for 27.2\,d with a sample mass of 174.1\,g. Upper limits are reported at 90\% C.L.\ ($k=1.645$); measurement uncertainties at 68\% C.L.\ ($k=1$). For the $^{232}$Th chain, $^{228}$Ra and $^{228}$Th are detected in EJ-560 at $(1.7\pm0.7)$\,mBq/kg and $(1.8\pm0.5)$\,mBq/kg respectively. Both Sylgard\,184 and EGel\,3002\,A/B yield no detectable $^{232}$Th-chain activity, with upper limits at the sub-mBq/kg level for the \texttt{GePaolo} measurement. For the $^{238}$U chain, $^{226}$Ra is detected in Sylgard\,184 and EGel\,3002\,A/B at $(2.4\pm0.8)$\,mBq/kg and $(2.2\pm0.7)$\,mBq/kg respectively, while remaining below the detection threshold in EJ-560 ($<$1.3\,mBq/kg). The upper parts of the $^{238}$U chain, probed via $^{234}$Th and $^{234m}$Pa, are undetected in all three materials. $^{40}$K is detected in all samples, at $(90\pm10)$\,mBq/kg for Sylgard\,184, $(50\pm8)$\,mBq/kg for EGel\,3002\,A/B, and $(120\pm20)$\,mBq/kg for EJ-560. No $^{137}$Cs contamination is found above the respective detection thresholds in any sample. The HPGe results for EJ-560 are complemented by HR-ICP-MS measurements performed on a 2.57\,g polymer foil sample, following the ashing and dissolution procedure described in Section~\ref{subsec:ICP-MS_Methodology}. Approximately 40\% of the sample mass remained unburned after ashing; accounting for this unburned fraction and assuming sample homogeneity, the estimated bulk concentrations are $(640\pm190)$\,ppt for $^{232}$Th and $(240\pm70)$\,ppt for $^{238}$U, corresponding to activities of approximately 2.6\,mBq/kg and 3.0\,mBq/kg respectively.

\begin{table}[H]
\centering
\small
\renewcommand{\arraystretch}{1.4}
\caption{Screening results for the three optical pad materials. HPGe activities
are in mBq/kg; upper limits at 90\% C.L.\ ($k=1.645$), uncertainties at 68\%
C.L.\ ($k=1$). Chain activities are inferred from the indicated progeny, given in parentheses.
ICP-MS concentrations for EJ-560 originally reported in ppt; equivalent activity
is stated in mBq/kg for comparison}

\noindent\makebox[\textwidth][c]{%
\begin{tabular}{l c c c c}
\toprule
\multirow{3}{*}{\textbf{Nuclide}} &
\multicolumn{3}{c}{\textbf{HPGe (mBq/kg)}} &
\multirow{2}{*}{\makecell{\textbf{ICP-MS}\\\textbf{(ppt)}}} \\
\cmidrule(lr){2-4}
& \textbf{Sylgard 184} & \textbf{EGel 3002\,A/B} & \textbf{EJ-560} & \\
& \texttt{GsOr} & \texttt{GePaolo} & \texttt{GeMPI3} & \\
\midrule
\multicolumn{5}{l}{\textit{$^{232}$Th chain}} \\
$^{228}$Ra\,($^{228}$Ac)
    & $<2.9$      & $<1.6$      & $1.7\pm0.7$
    & \multirow{2}{*}{\makecell{$640\pm190$\\\,$(2.6\pm0.8$ mBq/kg)}} \\
$^{228}$Th\,($^{212}$Pb, $^{212}$Bi, $^{208}$Tl)
    & $<3.0$      & $<2.2$      & $1.8\pm0.5$ & \\
\midrule
\multicolumn{5}{l}{\textit{$^{238}$U chain}} \\
$^{226}$Ra\,($^{214}$Pb, $^{214}$Bi)
    & $2.4\pm0.8$ & $2.2\pm0.7$ & $<1.3$
    & \multirow{3}{*}{\makecell{$240\pm70$\\\,$(3.0\pm0.9$ mBq/kg)}} \\
$^{234}$Th
    & $<55$       & $<47$       & $<17$       & \\
$^{234m}$Pa
    & $<55$       & $<36$       & $<38$       & \\
\midrule
$^{235}$U
    & $<5.1$      & $<1.3$      & $<0.58$     & --- \\
\midrule
\multicolumn{5}{l}{\textit{Other}} \\
$^{40}$K
    & $90\pm10$   & $50\pm8$    & $120\pm20$  & --- \\
$^{137}$Cs
    & $<0.94$     & $<0.76$     & $<0.30$     & --- \\
\bottomrule
\end{tabular}%
}
\label{tab:screening_results}
\end{table}

\section{Discussion}
\label{sec:discussion}

\subsection{Optical performance}

The  Sylgard\,184 based samples show a non-monotonic response in terms of absorption coefficient increasing catalyst fraction: $\mu$ falls from $0.277\pm0.016$~mm\textsuperscript{-1} at 10:1 to a minimum of $0.097\pm0.015$~mm\textsuperscript{-1} at 10:2, then rises sharply to $0.504\pm0.028$~mm\textsuperscript{-1} at 10:3 (RT values). This behavior may suggest that casting quality and structural imperfections—such as impurities and micro-bubbles—play a more significant role in determining the final transmittance and absorption coefficient values rather than the existence of an optimal catalyst ratio window. In another hand, curing temperature's effect on Sylgard\,184 is clearly ratio-dependent: Warming from RT to 40$^\circ$C significantly lowers $\mu$ at 10:1 (from $0.277\pm0.016$ to $0.178\pm0.013$~mm\textsuperscript{-1}) but raises it at 10:3 (from $0.504\pm0.028$ to $0.588\pm0.024$~mm\textsuperscript{-1}), both shifts exceeding their combined uncertainties. At 10:2, the RT and 40$^\circ$C values ($0.097\pm0.015$ and $0.093\pm0.014$~mm\textsuperscript{-1}) overlap within uncertainty, showing no resolvable temperature effect at this ratio.

\smallskip

For EGel\,3002\,A/B, the large gap between sample \#7 ($\mu = 0.439$~mm\textsuperscript{-1}, RT) and sample \#8 ($\mu = 0.125\pm0.015$~mm\textsuperscript{-1}, 40$^\circ$C) at the same 2:1 B-to-A ratio points to a critical role for thermal activation, likely driving the crosslinking reaction to higher conversion --- though the RT cured pad value's reliability is itself limited by that sample's surface inconsistency (see Table \ref{tab:thick_transmittance}).

\smallskip

The hybrid A:CataSYL configuration --- in which the EGel A-component serves as the optical matrix while CataSYL drives crosslinking --- decouples two functional roles that are inseparable in the standard EGel and Sylgard\,184 formulations, in principle allowing mechanical and optical properties to be optimized independently. At room temperature (sample~\#9), it reaches $\mu = 0.106\pm0.014$~mm\textsuperscript{-1}, the third-closest value to the EJ-560 reference, behind both Sylgard\,184 10:2 variants. The rise to $\mu = 0.275\pm0.027$~mm\textsuperscript{-1} at 40$^\circ$C (sample~\#10) is a statistically significant degradation; in opposition to the parent EGel system, where warming lowered $\mu$. This inversion suggests that EGel A-component/CataSYL compatibility is sensitive to curing conditions in a way neither native formulation shows, though the microscopic origin cannot be determined from absorption data alone. Characterizing the cured network structure, e.g.\ via gel-fraction analysis or optical microscopy, would help test further hypotheses and possibly identify an optimal curing protocol.

Overall, four of the locally produced pads approach the EJ-560 reference: Sylgard\,184 at 10:2 (samples \#3 and \#4), EGel\,3002\,A/B at 40$^\circ$C (sample \#8), and the A:CataSYL hybrid at RT (sample \#9). From their absorption coefficients, we estimate the thickness each pad would need to match EJ-560's transmittance: approximately 0.80~mm for Syl~10:2@RT, 0.84~mm for Syl~10:2@40\textdegree C, 0.62~mm for EGEL~2B:A@40\textdegree C, and 0.74~mm for A:CataSYL~10:1@RT. These latter results show that in-house production of optical coupling pads with performance comparable to the commercial material is achievable using precision casting or more advanced techniques such as Mayer rod coating for low but uniform thickness of samples.

\subsection{Radiopurity}

The contamination profiles of the three materials are complementary, and
their interpretation for the SABRE-North background budget mainly requires considering isotopic chains
emissions producing signals (tails included) in the 2 to 6 keV energy region. The detection of both $^{228}$Ra and $^{228}$Th at comparable levels of $(1.7\pm0.7)$ and $(1.8\pm0.5)$\,mBq/kg in EJ-560 indicates secular equilibrium within the lower thorium chain. The ICP-MS bulk concentration of $(640\pm190)$\,ppt for $^{232}$Th (corresponding to $\sim$2.6\,mBq/kg) is broadly consistent with this, confirming that the $^{232}$Th contamination is distributed throughout the material matrix rather than being a surface artifact. The absence of any detectable $^{232}$Th-chain activity in both locally produced compounds, with sub-mBq/kg upper limits, is therefore a genuine advantage of those materials over EJ-560.

The detection of $^{226}$Ra in Sylgard\,184 and EGel\,3002\,A/B at $(2.4\pm0.8)$ and $(2.2\pm0.7)$\,mBq/kg respectively, without any accompanying signal from the upper $^{238}$U chain (via $^{234}$Th and $^{234m}$Pa), points to a disruption of secular equilibrium between uranium and radium. This pattern is consistent with chemical fractionation during industrial manufacturing: radium, being alkaline-earth-like, can concentrate preferentially relative to uranium in processes involving precipitation or adsorption steps. Alternatively, since the $^{226}$Ra is estimated from its daughters $^{214}$Pb and $^{214}$Bi, it may reflect a specific $^{222}$Rn contamination through trapped micro-air bubbles introduced during synthesis~\cite{cooley2013}\cite{WESTERDALE2017}.

The detected $^{226}$Ra itself drives $^{210}$Pb —a major background contributor in the 2--6\,keV ROI- to $\sim$2.3\,mBq/kg as the sub-chain equilibrates, regardless of the state of the chain above it. However, quantifying the full $^{238}$U and $^{232}$Th chain activities remains important independently: $\alpha$-emitting members of both chains can contribute to the detector background in the ROI through surface interactions and quenching effects, and their contribution cannot be bounded from the $^{226}$Ra measurement alone, because the HPGe detection thresholds for the upper-chain proxies $^{234}$Th and $^{234m}$Pa ($\sim$47--55\,mBq/kg) exceed the detected $^{226}$Ra activity by more than an order of magnitude — a consequence of the weak and low-energy gamma emission of $^{234}$Th ($E_\gamma = 63$\,keV, branching ratio $\sim$4\%) — leaving open the possibility of a substantially larger full-chain activity that is simply invisible to the present measurement. ICP-MS measurements of $^{232}$Th and $^{238}$U in the locally produced materials, analogous to those performed for EJ-560, are therefore a priority for future work. If the measured $^{238}$U concentration is consistent with the observed $^{226}$Ra activity, this would indicate that the $^{238}$U decay chain is close to secular equilibrium and would consequently constrain the expected $^{210}$Pb activity. Conversely, if the $^{238}$U concentration is found to be significantly higher than implied by the $^{226}$Ra activity, it would suggest that the decay chain is out of equilibrium and that the $^{210}$Pb contribution to the ROI has been underestimated.

Regarding $^{40}$K — the other primary background contributor in the 2--6\,keV ROI — EJ-560 exhibits the highest activity at $(120\pm20)$\,mBq/kg, exceeding both Sylgard\,184 at $(90\pm10)$\,mBq/kg and EGel\,3002\,A/B at $(50\pm8)$\,mBq/kg. This finding runs counter to the assumption that a commercial, purpose-designed low-background pad would necessarily outperform locally produced alternatives on all contamination metrics.

An important caveat applies to all locally produced material measurements: screening was performed on uncured liquid mixtures, not on finished pads. While the curing reaction itself is not expected to introduce radioactive contamination, the handling steps involved in pad production — mold contact, laboratory environment, and uncontrolled ambient conditions — could in principle alter the surface or bulk contamination of the final product. A dedicated HPGe screening of fully cured pads produced under clean-room conditions is therefore necessary to confirm that the liquid-phase results are representative of the material as it would be deployed in the detector.

As of this work, no single material is unambiguously superior across all performance criteria examined here. EJ-560 presents the lowest absorption coefficient and a well-characterized contamination profile, but carries detectable Th-chain activity and the highest $^{40}$K level of the three candidates. The Sylgard\,184 10:2 samples and the hybrid A:CataSYL pads approach EJ-560's optical performance with a possibility of reaching it using thinner samples. They exhibit no detectable Th-chain contamination, and show lower $^{40}$K levels, but present $^{226}$Ra activity for which chain context requires clarification. EGel\,3002\,A/B cured at 40$^\circ$C combines promising absorption coefficient with the lowest $^{40}$K activity measured in this study, though the same $^{226}$Ra caveat applies. Translating these measured activities into a detector-level background contribution requires propagating them through a full SABRE-North geometry simulation to estimate their rate in the 2--6\,keV ROI, which lies beyond the scope of this work, but represents the natural next step once the outstanding measurements described below are completed.

\section{Conclusions}
\label{sec:conclusions}

This work presents a systematic optical and radiopurity characterization of three optical coupling pad materials — EJ-560, Sylgard\,184, and EGel\,3002\,A/B — as candidates for the SABRE-North dark matter detector. Ten locally produced pads were fabricated with varying catalyst ratios and curing conditions, and evaluated against the EJ-560 commercial reference through UV/Vis/NIR transmittance spectrophotometry and HPGe $\gamma$-ray spectrometry at the STELLA facility, complemented by further HR-ICP-MS screening on the EJ-560.

Four of the locally produced optical pads approach the absorption coefficient of the EJ-560 reference material, though none are compatible with it within the 2\% relative uncertainty on transmittance. The best performers, Sylgard,184 at 10:2 (both curing temperatures), still show a resolvable excess absorption relative to EJ-560. This nonetheless demonstrates that in-house fabrication of optical pads approaching commercial-grade performance is feasible at significantly lower cost than the commercial alternative. The A:CataSYL hybrid pads, which decouples optical and mechanical formulation roles, also remains a promising avenue for future research  despite its curing-condition sensitivity. Also, considering that the absorption coefficients of Syl 10:2 @RT, Syl 10:2 @40°C, EGEL 2B:A @40°C, and A:CataSYL 10:1 @RT are relatively close to that of the EJ-560 reference pad, it can be concluded that transmittance values of $\sim$92\% can be achieved by optimizing the pad thickness. Furthermore, it is likely that a significant fraction of the non-transmitted radiation is not due to absorption, but rather to scattering effects caused by surface imperfections and/or internal defects within the polymer layers, such as the presence of micro-bubbles. Therefore, the selection of the most suitable material should primarily consider the radioactive emission properties of the pads, in particular the presence of contaminants that could negatively affect the SABRE-North experiment. Since the optical performance of the home-made pads can still be improved through optimization of the fabrication process, material selection should be mainly driven by radiopurity requirements.

\medskip

On the radiopurity side, the three materials exhibit complementary contamination signatures that prevent a straightforward ranking. EJ-560 is the only material with detectable Th-chain activity and carries the highest $^{40}$K concentration among the three candidates, whereas the locally produced compounds -or at least their mother solutions- show detectable $^{226}$Ra activity whose relationship to the broader $^{238}$U chain remains to be established, alongside lower or comparable $^{40}$K levels. The production flexibility of Sylgard\,184 and the hybrid A:CataSYL system, combined with their promising optical and radiopurity profiles, identifies these two routes as priority candidates for the next characterization phase. 

The measurements outstanding — ICP-MS screening of the locally produced materials, HPGe screening of fully cured pads under clean-room conditions, refractive index characterization, and direct light-yield testing in a NaI(Tl)--PMT assembly — are required to resolve the remaining ambiguities and provide a well-founded recommendation for the SABRE-North detector.

\section*{Acknowledgments}
This work was supported by the European Union – Next Generation EU under the PRIN 2022 PNRR program (D.D. 1409, September 14, 2022), Project ID P2022L9XZC - “Production of Ultra-pure NaI detecTOrs by zone refining (PUNTO)”; by the National Science Centre, Poland [Grant No. 2025/59/B/ST2/02359]; and by the research project “MAD – La metamorfosi Additiva del Design" – cod. ARS01\_00717 (CUP B82F20000690005). The authors gratefully acknowledge the Chemistry Laboratory and the STELLA facility of the Laboratori Nazionali del Gran Sasso (LNGS) for providing the technical support essential to this research.


\bibliographystyle{JHEP}
\bibliography{biblio}


\end{document}